\documentclass{article}
\usepackage{palatino}
\usepackage{graphicx} 
\usepackage{fullpage}
\usepackage{natbib}
\usepackage{authblk}
\usepackage{coffeestains}
\usepackage{subcaption}
\usepackage{url}
\usepackage{lineno}

\title{Global Prevalence of Anti-similar Earthquakes \\ at Intermediate Depth}
\author[1*]{Albert Leonardo Aguilar Suarez}
\author[1]{Gregory Beroza}
\author[1]{William Ellsworth}
\author[2]{Patricia Pedraza}
\author[3]{Germán Prieto}
\affil[1]{Stanford University}
\affil[2]{Servicio Geológico Colombiano}
\affil[3]{Universidad Nacional de Colombia}
\affil[*]{corresponding author, aguilars@stanford.edu}

\date{August 2026}

\begin{document}
\maketitle

\section{Abstract}

Anti-similar earthquakes, those with nearly identical but flipped waveforms, remain an uncommon observation. We document three locations where these events occur at intermediate depth. The first of these is Colombia, including the well-established source of anti-similar events in the Bucaramanga Nest, and new reports in other intermediate depth seismicity clusters in the Wadati-Benioff zones of Colombia. The second is the Alaska subduction zone, and the third is the Vrancea seismic nest in Romania. These new observations, combined with reports from Japan and Chile, provide mounting evidence that anti-similar earthquake sequences might be a common or characteristic process of intermediate-depth seismicity. Either conjugate faulting promoted by eclogitization induced stresses or sub-parallel faulting can explain their occurrence at these depths. Careful examination of seismicity in other productive intermediate depth seismicity zones might reveal an even more widespread geography for anti-similar earthquakes and place tighter constraints on their underlying mechanism.



\section{Introduction}

Intermediate depth earthquakes, defined as those with hypocentral depths between 70 and 300 km, and deep earthquakes  defined as those with hypocentral depths greater than 300 km, comprise 25\% of observed global seismicity.  They are recognized for their highly clustered nature, heterogeneous focal mechanisms, and high frequency waveforms \citep{Green1995,Zhan2020}. The mechanisms behind intermediate depth earthquakes are debated, due in part to limitations in observations. In contrast to shallow fault slip, and shallow seismicity, which can be observed through seismological, geodetic, satellite, borehole, and strainmeter data, insights into intermediate depth seismicity are almost entirely restricted to the information carried by seismic waves, and mechanics based on theory or laboratory experiments. Although the hazard posed by intermediate depth seismicity is less than that from shallow seismicity, it is nevertheless  significant in active subduction zones such as Alaska, Colombia, and relic subduction zones such as Romania, and Central Asia \citep{Prieto2012}. All of these locations can host earthquakes approaching magnitude 8. For instance, the deadliest earthquake in Chile was the 1939 Chillán earthquake Ms 7.8, which ruptured the slab at a depth of 100 km causing 26,000 casualties, far more than the 1,600 fatalities caused by the M$_w$ 9.5 1960 Valdivia earthquake, the largest instrumentally recorded earthquake \citep{BECK1998115}. In Colombia, the 2026 San José del Palmar Mw 7.4 earthquake left hundreds of casualties, and thousands of homes and buildings badly damaged. In Romania, the 1940 and 1977 Vrancea earthquakes are the most destructive and deadly in the country's history \citep{FUCHS1979225}. The processes proposed to be the cause of earthquakes at intermediate depth include thermal shear runaway instability \citep{Poli2016}, dehydration embrittlement \citep{Jung2004}, dehydration driven stress transfer \citep{Ferrand2017}, and lithospheric delamination \citep{Perez-Forero2023}.

Anti-similar earthquakes, or anti-repeating earthquakes are events that have high waveform similarity, but with the polarity reversed on all components (Figure \ref{fig:alaska_wav}).  Observations of such events have emerged recently in a variety of contexts, including volcanic, induced, and tectonic seismicity \citep{Cesca2024}. The body of research around anti-similar earthquakes is growing as well. \citep{eaton1970} is one of the earliest references to earthquakes of this type during the aftershocks of the 1966 Parkfield earthquake, where earthquakes with nearly opposite focal mechanisms occurred in close proximity. \citep{zobackberoza93} documented opposite focal mechanisms for aftershocks of the 1989 Loma Prieta earthquake in California. \citep{Hauksson2005} reported the occurrence of negative waveform correlations in Southern California as well. In subduction zones, shallow dynamic overshoot is thought to have produced earthquakes with slip in opposite directions, and anti-correlated waveforms in the rupture area of the 2011 Tohoku earthquake \citep{ide2011}, the 2014 Iquique earthquake \citep{cesca_2016}, and the 2025 Kamchatka earthquake \citep{Yagi_Fukahata_Okuwaki_Takagawa_Toda_2025}, all low angle normal faulting earthquakes after megathrust events. 

In other tectonic settings, analogous observations have been described, including the 2014 Long Valley caldera swarm \citep{Shelly2016} in California. \citep{Trugman2020} reported sequences of anti-correlated earthquakes in orthogonal fault strands during the 2019 Ridgecrest earthquake sequence in California. \citep{Lu2025} found 37 pairs of anti-similar earthquakes in Sierraville California, most likely in the mantle. \citep{lu2026} also reported anti-similar events in the Abu magmatic sequence of 2025 in Japan. \citep{Cesca2024} reviewed reports of anti-repeating earthquakes, and  performed a global search for anti-similar earthquakes that shows the extent of these events for moderate to large magnitudes \citep{cesca2026}. It is clear that in many of these observations earthquakes with anti-correlated waveforms occur on separate fault planes, which might be the most common case, where modeling indicates that even on different planes, earthquakes with orientation differences up to 20 degrees could produce the anti-similar signature \citep{shearer2024}. There are also scenarios that point to repeated rupture of the same fault, but in opposite directions, mainly in volcanic settings, for example in Mayotte \citep{Cesca2020}, and  also for trapdoor faulting \citep{Snadanbata_2022}, and ring faulting \citep{shuler_2013}.

At intermediate depth, the report of anti-similar earthquakes in \cite{Prieto2012} in the Bucaramanga Nest in eastern Colombia is to our knowledge the earliest account of this phenomenon. Subsequently, \cite{nakajima2013} reported similar observations in Japan, and offered an explanation for these events in which a compact cluster of tensional earthquakes overlying a cluster of compressional earthquakes indicate stresses induced by eclogitization, while also posing the scenario of sub parallel faults. \cite{plourde_2019} reported several anti-similar events for a cluster in southeast Alaska at 125 km depth, attributing them to eclogitization of gabbro. More recently \cite{folesky_anti} reported similar observations in Northern Chile. Here, we present new observations of anti-similar earthquakes that we discovered through systematic similarity search at a number of locations that are highly productive in intermediate depth seismicity. Colombia's intermediate depth seismicity belts yielded the highest number of sequences containing anti-similar events, between 100 and 150 km depth. We further document intermediate depth anti-similar events in Alaska. Finally, we include a single observation in the Vrancea intermediate depth seismicity nest in Romania. We describe these events as "anti-similar", because we hypothesize that they are occurring on different faults, and reserve anti-repeating for repeated slip on the same fault patch in opposite directions. Once their generative mechanism is determined, the observations summarized here have the potential to inform the underlying mechanisms of intermediate depth seismicity.

\section{Methods}

Our initial objective was to enhance seismicity catalogs in locations with intense intermediate depth seismicity, using template matching. For the purpose of this report, we focus on the highly similar and highly anti-similar detections. We found anti-similar earthquakes in Colombia, Alaska, and Romania (Figure \ref{fig:world}). In all cases, we performed multi-station template matching over years of continuous data. We formed the templates from continuous data band-pass filtered between 1 and 15 Hz, and scanned them on continuous data using EQcorrscan \citep{eqcorrscan}. We selected as candidates for similar and anti-similar earthquakes those with a network-averaged cross correlation value higher than 0.8 or lower than -0.8. Some of the events are present in the routine earthquake catalogs while others are newly detected. For the newly detected events we estimated local magnitudes using amplitude ratios with respect to the template that detected them. We found a large number of similar earthquakes, with highly correlated waveforms, very close locations, and likely similar focal mechanisms. We found a smaller set of anti-similar earthquakes, implying very close locations, and opposite focal mechanisms.

\section{Results}

\begin{figure}[h]

\begin{subfigure}{0.5\textwidth}
\includegraphics[trim={2.75cm 0.1cm 2cm 0.1cm},clip,width=\linewidth]{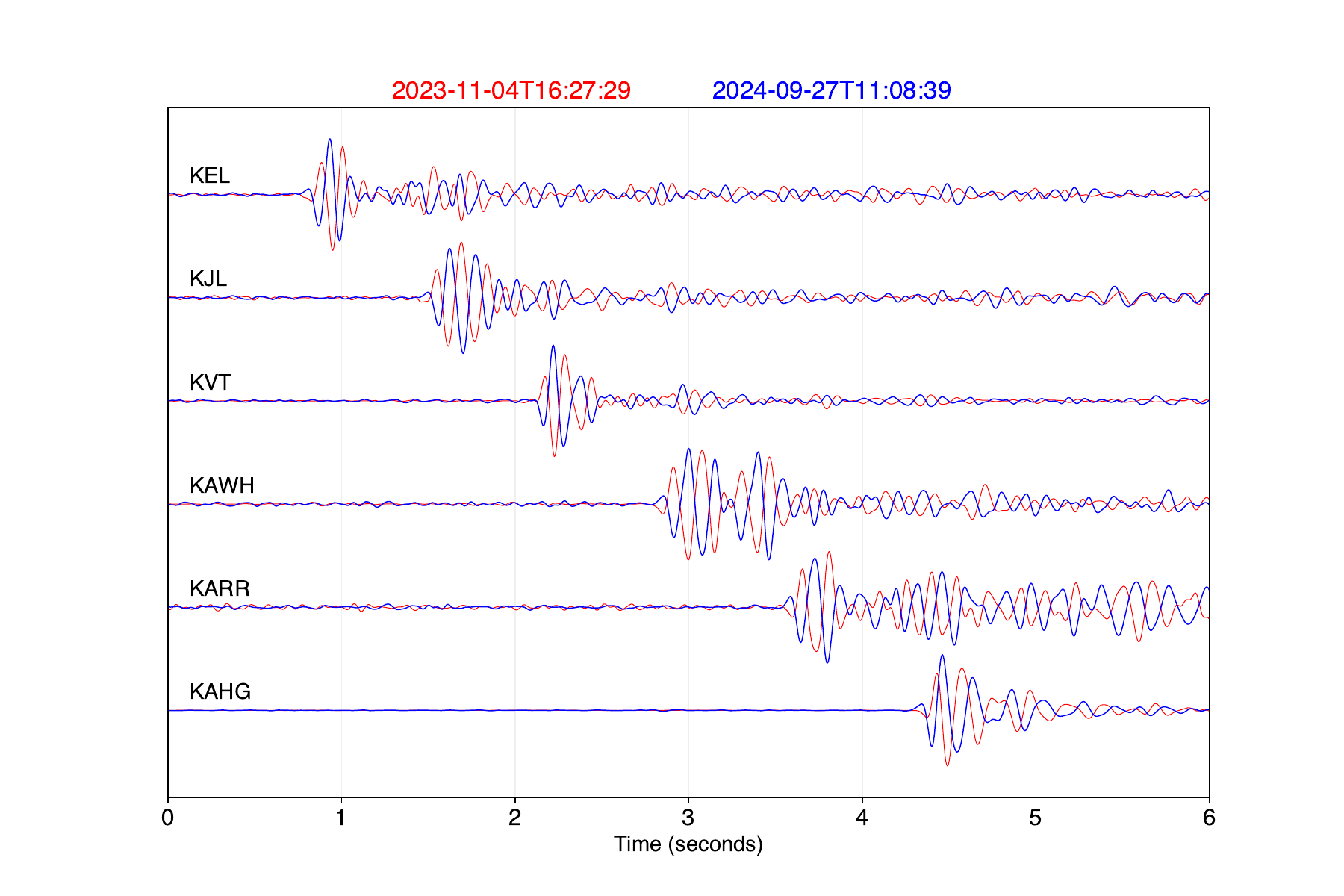} 
\caption{Original waveforms}
\label{fig:subim1}
\end{subfigure}
\begin{subfigure}{0.5\textwidth}
\includegraphics[trim={2.75cm 0.1cm 2cm 0.1cm},clip,width=\linewidth]{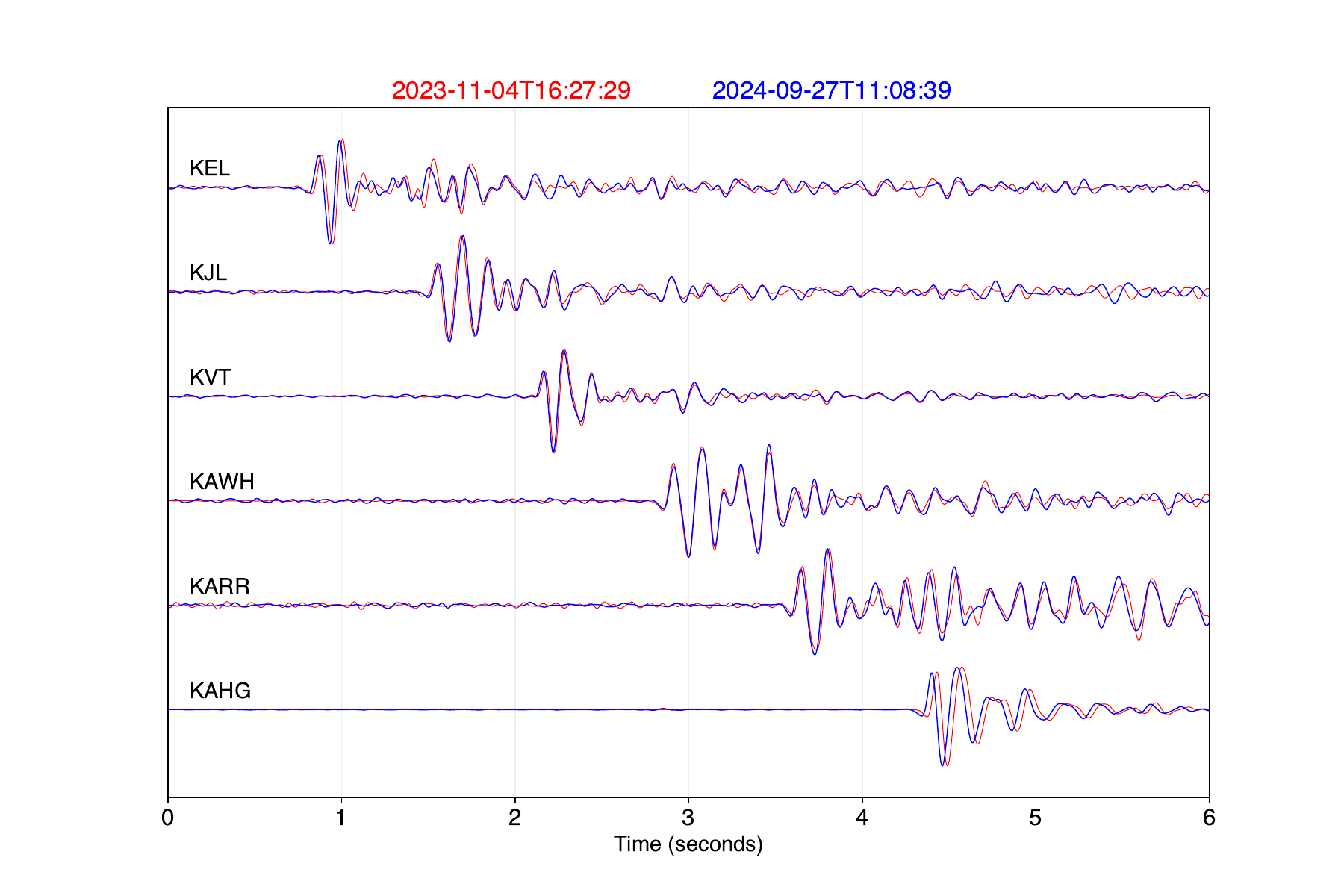}
\caption{Flipped event 2}
\label{fig:subim2}
\end{subfigure}
\caption{Antisimilar earthquakes in the Alaska subduction zone. Waveforms are highpass filtered at 5 Hz. Vertical channel only. Stations are sorted by distance to the epicenter.}
\label{fig:alaska_wav}
\end{figure}

\begin{figure}[h!]
    \centering
    \includegraphics[trim={1.5cm 1.5cm 2cm 8cm},clip,width=0.95\linewidth]{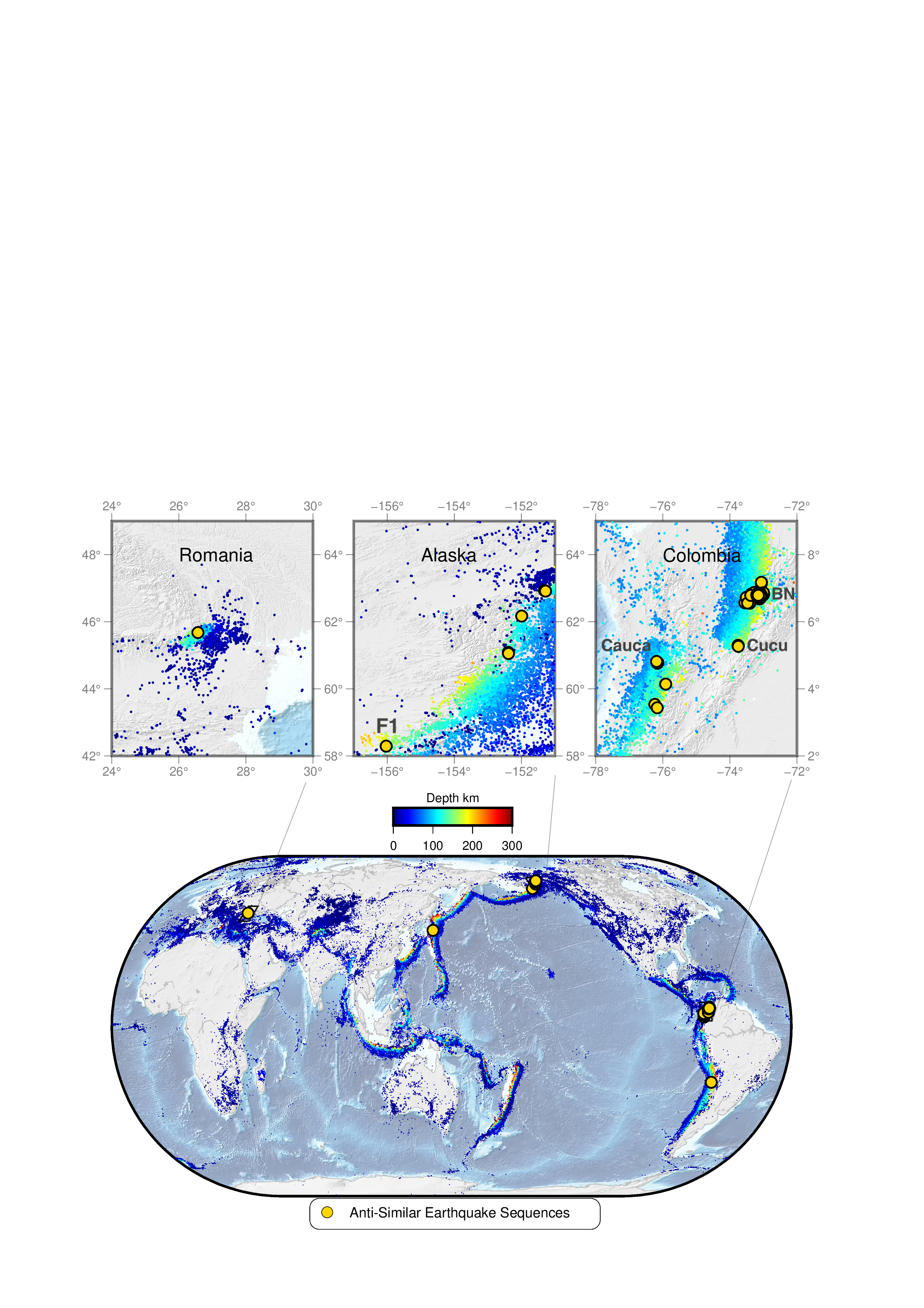}
    \caption{Locations of recently reported intermediate depth anti-similar earthquake sequences. The locations for the Japanese sequences are taken from \citep{nakajima2013}, the ones for Chile from \citep{folesky_anti}. Global seismicity from the ISC catalog. In the Colombia panel we removed shallow seismicity for ease of visualization. BN stands for Bucaramanga Nest, Cucu for Cucunubá cluster, and Cauca indicates the Cauca cluster in the Cauca WBZ. In the Alaska panel F1 indicates the location of the anti-similar pair of earthquakes shown in Figure \ref{fig:alaska_wav}. }
    \label{fig:world}
\end{figure}

Our similarity search of earthquakes in three regions is based solely on waveform cross correlation over multiple stations at high frequencies. An example of a detection over multiple stations in Alaska is shown in Figure \ref{fig:alaska_wav}. In the right panel the waveforms have been flipped to better visualize the similarity. Notice the slight shift at the top and bottom stations, whereas the alignment is near perfect in the middle stations. This is the signature of close, but not identical hypocenters for these events. Next, we detail our findings for three regions: Colombia, Alaska, and Romania, which are ordered by decreasing productivity and shown in Figure \ref{fig:world}.

\vspace{5mm}

\subsection{Colombia}

Colombia is particularly productive in intermediate depth seismicity, with 60 \% of seismicity reported by the Servicio Geológico Colombiano (SGC) occurring below 70 km depth, distributed in two distinct Wadati-Benioff zones (WBZ), the Bucaramanga and the Cauca, as a consequence of the subduction of the Nazca and Caribbean plates under South America \citep{Vargas2020}. These two WBZs are offset 250 km by the Caldas Tear in central Colombia \citep{Vargas2013,Vargas2020}. The tear separates contrasting subduction styles, featuring a flat slab in the north \citep{Wagner2017} and a normally dipping Cauca slab in the south \citep{chang2017} that is highly influenced by the interaction with the Panama-Chocó block \citep{wagnerpanama_2025,bishop2025} (Figure \ref{fig:world}). Events notable for their magnitudes and effects include the Bucaramanga Nest Mw 6.7 in 1967 \citep{ramirez1974}, and more recently a Mw 6.3 in 2015 \citep{Poli2016}. In the Cauca WBZ, the 2012 Mw 7.2 La Vega intermediate-depth earthquake, and the very recent 2026 San José del Palmar Mw 7.4 are the largest recorded in the last decades.

Anti-similar earthquakes at intermediate depth were originally reported by \cite{Prieto2012} in the Bucaramanga Nest in eastern Colombia, an intense and compact volume of seismicity at a depth of approximately 150 km that produces a magnitude 4 and above event every 11 days on average per the SGC catalog and has long been recognized as a hotspot of persistent seismicity \citep{Frohlich1995,ZARIFI2003237}. This nest is located on a kink of the Bucaramanga WBZ, that extends hundreds of kilometers both north and south in a continuous along strike continuity belt of seismicity (Figure \ref{fig:world}). 

\begin{figure}
    \begin{subfigure}{\textwidth}
        \includegraphics[trim={1.5cm 0cm 2cm 0cm},clip,width=\linewidth]{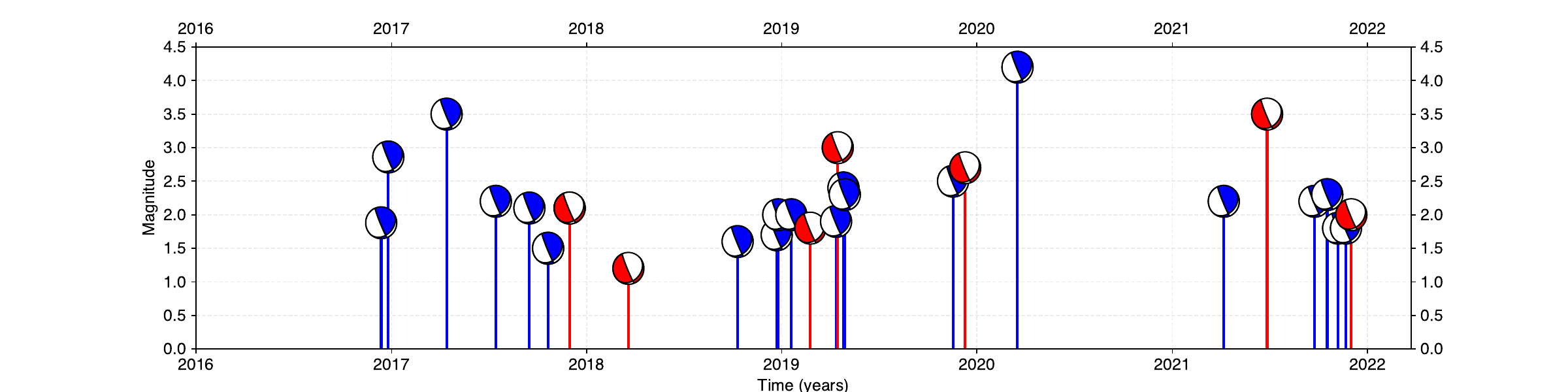}
    \end{subfigure}

    \begin{subfigure}{\textwidth}
        \includegraphics[trim={1.5cm 0cm 2cm 0cm},clip,width=\linewidth]{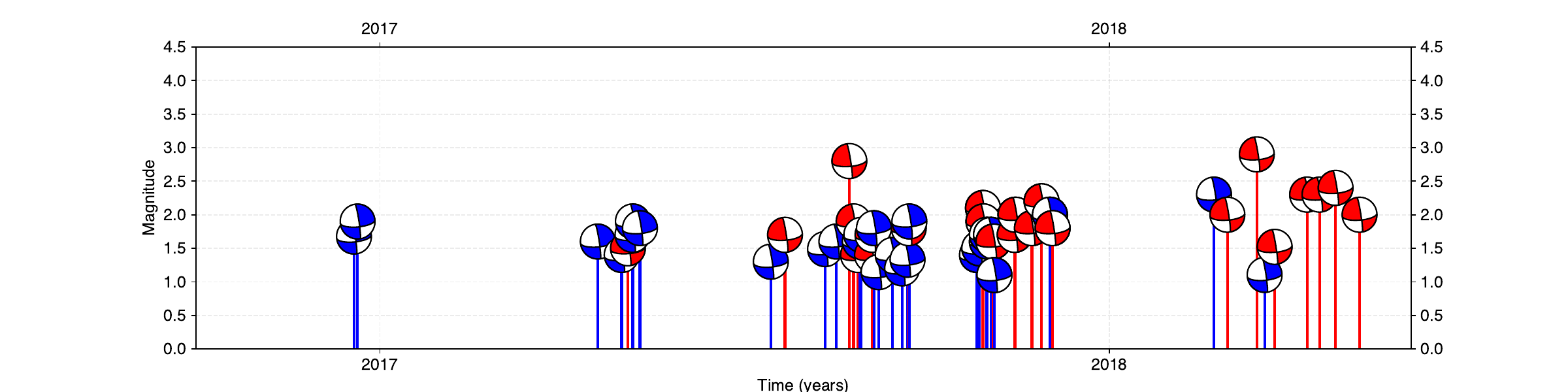}
    \end{subfigure}

    \begin{subfigure}{\textwidth}
        \includegraphics[trim={1.5cm 0cm 2cm 0cm},clip,width=\linewidth]{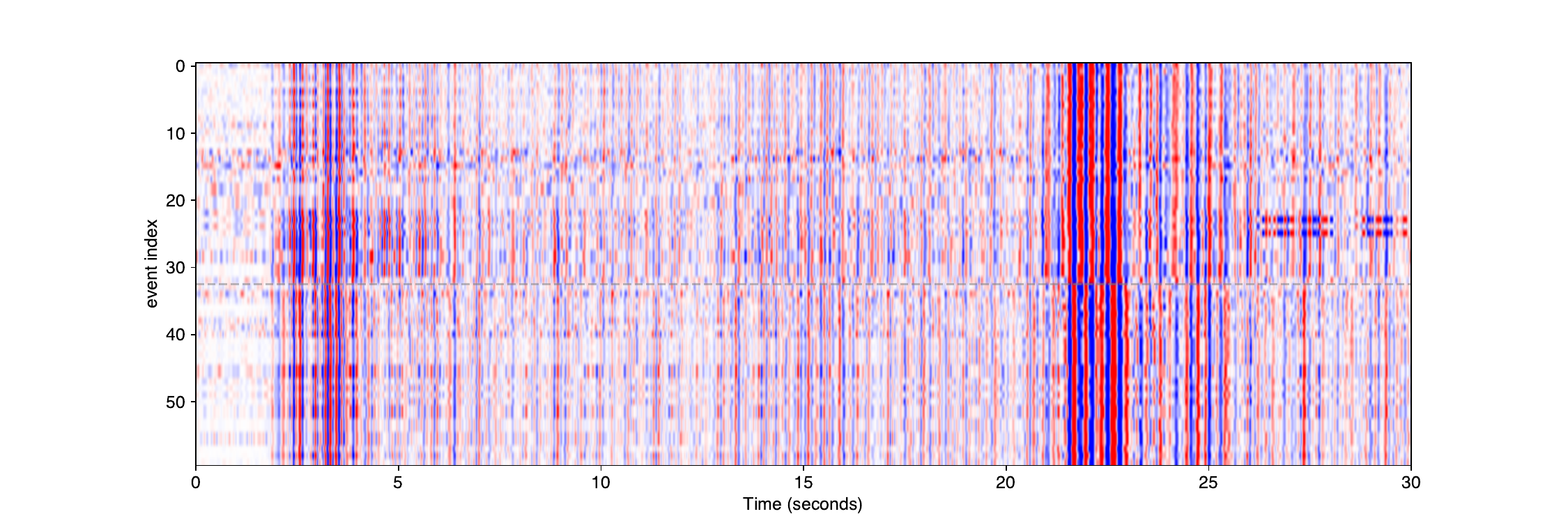}
    \end{subfigure}

    \caption{Anti-similar sequences in the Bucaramanga Nest. (Top) One of the longest-lived sequences we identified, active from late 2016 until late 2021. (Middle) One of the most well balanced sequences, with 34 and 31 end members, active from late 2016 until mid 2018. (Bottom) Waveforms for the sequence in the middle panel, as recorded by the vertical component of station CM.RUS, bandpass filtered 1-10 Hz. The waveforms for each polarity are separated by the gray dotted line at index 33, with the events from the blue focal mechanisms above the line, and the ones from the red focal mechanism below. A few of the 65 earthquakes in this sequence occurred when station CM.RUS was offline.}
    \label{fig:bmanga}
\end{figure}

We conducted a template matching search on continuous data from the SGC seismic network for the period 2012 to 2023. We created templates by revisiting known origin times from the SGC catalog and measuring the arrival times with a bespoke skynet model \citep{AguilarSuarez_Beroza_2025}, which was trained on Colombian data. For the events that are in the SGC catalog we use their reported magnitudes, and for newly found events we estimated their magnitudes using amplitude ratios with their detecting template. For several of the events in these sequences the SGC reported focal mechanisms \citep{dionicio_fm} which we also use. We identified sequences with anti-similar earthquakes in the Bucaramanga Nest, the most prolific source of anti-similar earthquakes, its vicinity, and also in the Cucunubá cluster at the southern terminus of the Bucaramanga WBZ. On the other side of the country, the Cauca WBZ also hosts several anti-similar sequences (Figure \ref{fig:world}). These sequences exhibit a range of temporal behaviors, including quasi-periodic recurrence to swarm-like bursts. We identified both short-lived (several weeks) and long-lived (several years) sequences with anti-similar events. The most prolific sequence has 196 members, and there are many isolated pairs. The time lags range from minutes to over a year. Figure \ref{fig:bmanga}(top) shows the temporal evolution for one of the sequences in the Bucaramanga Nest that was active for almost six years, with 29 occurrences, of which 7 have opposite polarity. The magnitude of the events associated with this sequence range between 1.1 and 4.2. One of the best well-balanced sequences, with nearly equal numbers of similar vs. anti-similar waveforms, 34 and 31, is shown in  Figure \ref{fig:bmanga} (middle). The behavior is not balanced in time, however, as the first half is dominated by the blue focal mechanisms, while the second half is dominated by the red focal mechanisms. The waveforms (bandpass filtered 1-15 Hz) for the events in this sequence at station CM.RUS are displayed in Figure \ref{fig:bmanga}(bottom), separating the similar and the anti-similar ones, displaying the clear wiggle by wiggle coincidence with the polarity flip. Overall, we identified 992 sequences with anti-similar events at this location, with 5,985 events. There is at least one anti-similar earthquake occurrence in 80 \% of the days we analyzed. Overall, 7 \% of the seismicity in the Bucaramanga Nest belongs to an anti-similar earthquake sequence during our analysis. These events are distributed between depths of 110 and 157 km, with a peak at 145 km depth.

The Cucunubá cluster is the second most prolific location in Colombia for anti-similar earthquakes \citep{prieto2026}. This cluster produces abundant magnitude 3 and above events, at a depth of 145 km, similar to that of the Bucaramanga Nest, and also sits near the deeper end of the southern terminus of Bucaramanga WBZ (Figure \ref{fig:bmanga}). We identified 31 anti-similar sequences in Cucunubá, which account for 730 earthquakes. 
On the Cauca segment, we identified 8 sequences with 20 total events: one quartet, two triplets and 5 pairs. These sequences are distributed in 3 groups (Figure \ref{fig:world}). The first one in the Cauca cluster \citep{chang2017,Chang2019}, and two other groups farther south (Figure \ref{fig:world}). The anti-similar sequences in the Cauca cluster \cite{chang2017} have the shallowest depth, with a few sequences between 98 and 108 km, in a seismicity structure that is nearly perpendicular to the slab seismicity \citep{bishop2025}. The two other groups  are closer to the termination of the slab seismicity between 135 and 145 km depth.
The commonality for the Colombian anti-similar sequence locations except the one in the Cauca cluster is their proximity to the deeper end of the WBZs, concentrated in the 140-150 km depth range. For several of these locations there are also highly heterogeneous focal mechanisms in close proximity \citep{Prieto2012,Chang2019,dionicio_fm}.


\subsection{Alaska}

The Alaska-Aleutian subduction zone where the Pacific plate is being subducted under the North American plate is well-known for large megathrust earthquakes, but it has substantial intermediate depth seismicity as well \citep{ruppert_2024}, with around 20 \% of seismicity reported taking place between 70 and 300 km depth, including one of the largest intermediate depth earthquakes instrumentally recorded, the 2014 Rat islands Mw 7.9 event \cite{ye2014}. The slab seismicity is highly segmented \citep{wei2020_alaska}, featuring  flat slab subduction in the eastern end \citep{bauer2014}, and double and triple seismic zones \citep{zanjani2022}. 
\cite{plourde_2019} reported anti-similar earthquakes for a cluster at 125 km depth, in the flat slab where the Yakutat terrain is being subducted. They report 7 events distributed in 2 anti-similar sequences, one of 4 events and one with 3 earthquakes, both sequences with one reverse polarity member. For earthquakes in this cluster, their moment tensor estimates are mainly of normal and reverse faulting, but the surrounding seismicity shows high variability.

We conducted template matching for the period 2019-2025 using 196 stations from the Alaska Geophysical Network \citep{https://doi.org/10.7914/sn/ak}, and Alaska Volcano Observatory \citep{av}, using the vertical channels only. We  used the events in USGS ComCat as templates. Figure \ref{fig:alaska_wav} shows an example of anti-similar events found in Alaska at 143 km depth, both $M_{L}$ 1.9, occurring about 10 months apart in 2023 and 2024. Figure \ref{fig:alaska_wav}(a) shows  the original polarities of the waveforms while Figure \ref{fig:alaska_wav}(b) shows the waveforms for the 2024 event flipped, such that the polarities are now the same. Notice that for the first and last stations there is a noticeable shift in the waveforms; whereas, the middle ones show a near perfect alignment. This indicates that the events that do not perfectly align were in slightly different locations. Overall, we found 30 occurrences of anti-similar earthquakes, grouped in 6 sequences. One of these sequences contains 13 events, three sequences contain 4 events each, and there is one triplet and one doublet. These sequences occurred at depths of 105, 109, 110, 112, 116, and 143 km under the Alaska range, close to the down dip termination of slab seismicity (Figure \ref{fig:world}). The events in these sequences range in magnitude from 1.1 to 2.3, which makes estimating focal mechanisms challenging. 

\begin{figure}
    \begin{subfigure}{\textwidth}
        \includegraphics[trim={1.5cm 0cm 2cm 0cm},clip,width=\linewidth]{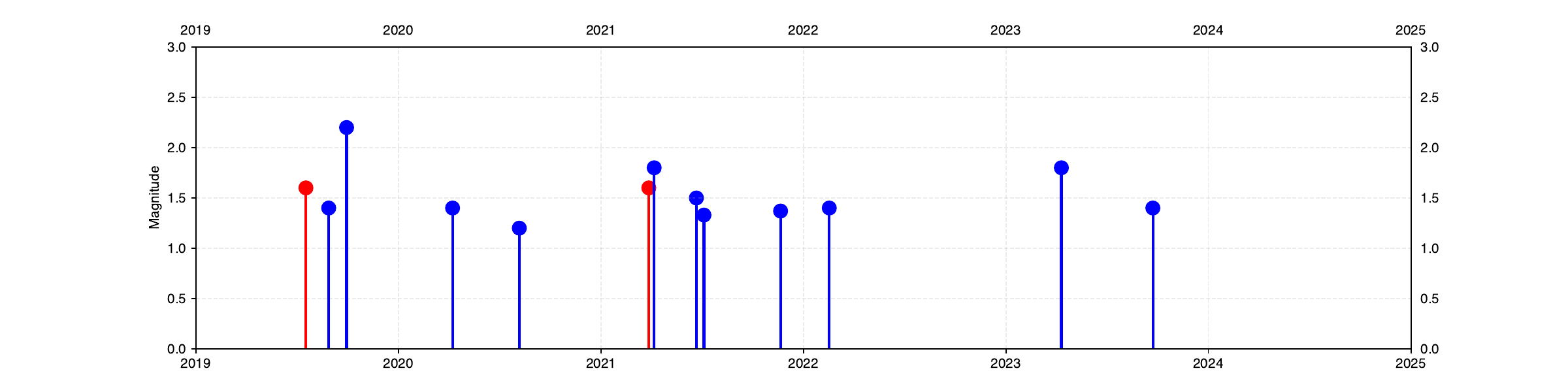}
    \end{subfigure}

    \begin{subfigure}{\textwidth}
        \includegraphics[trim={1.5cm 0cm 2cm 0cm},clip,width=\linewidth]{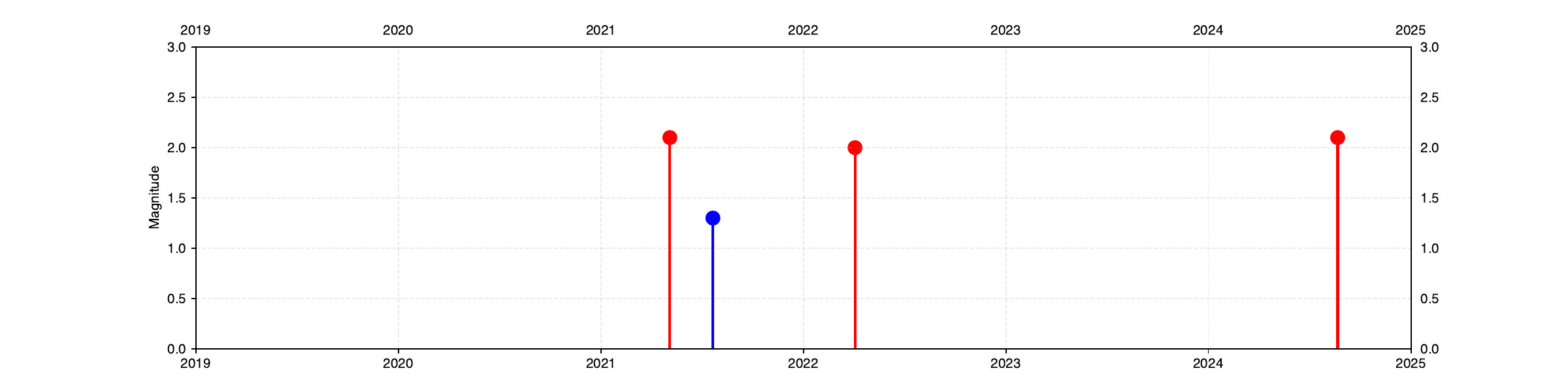}
    \end{subfigure}

    \begin{subfigure}{\textwidth}
        \includegraphics[trim={1.5cm 0cm 2cm 0cm},clip,width=\linewidth]{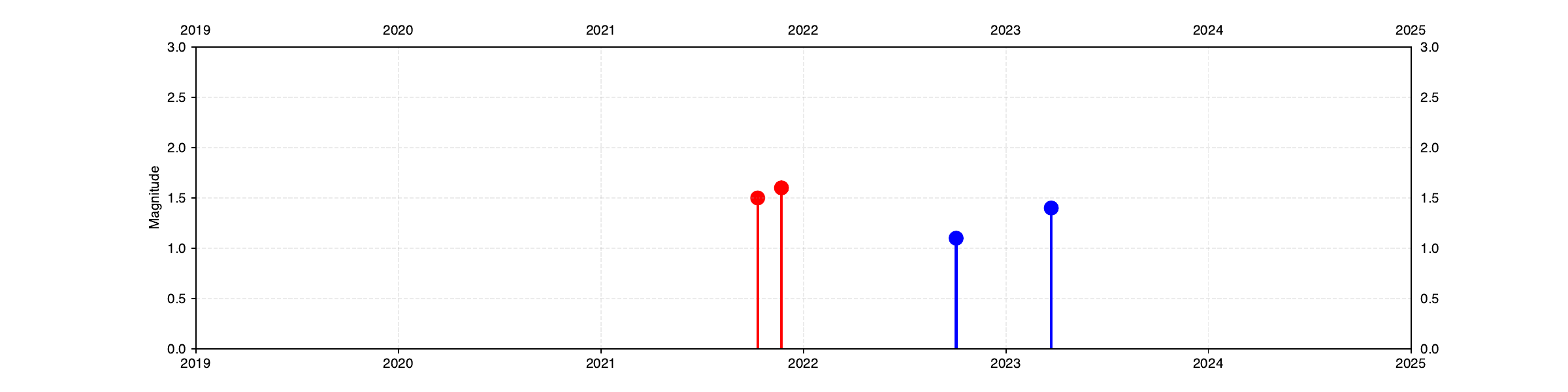}
    \end{subfigure}

    \begin{subfigure}{\textwidth}
        \includegraphics[trim={1.5cm 0cm 2cm 0cm},clip,width=\linewidth]{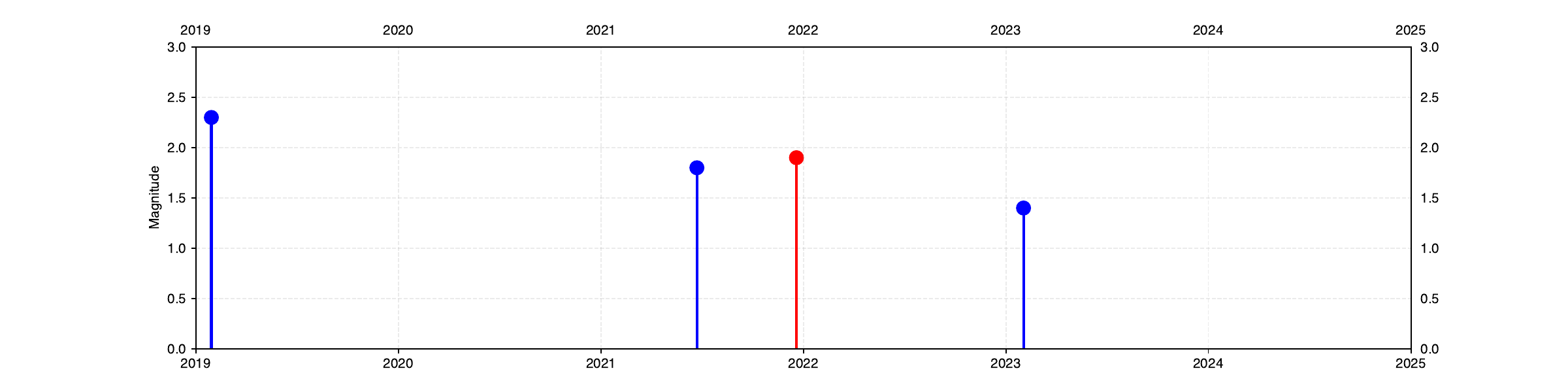}
    \end{subfigure}

    \caption{Four anti-similar earthquake sequences in Alaska.}
    \label{fig:alaska_sequences}
\end{figure}

Figure \ref{fig:alaska_sequences} shows the temporal evolution of the magnitudes for 4 of these sequences. The most productive sequence contains 13 events, 11 of one polarity and two of the opposite. This particular sequence is active for around 5 years, with earthquakes ranging in magnitude between 1.3 and 2.3. For this sequence the events closest in time are nearly 10 days apart in 2021, whereas the ones with the longest delay are 418 days apart, between the event in 2022 and the first one of 2023. The other three sequences in Figure \ref{fig:alaska_sequences} contain 4 events each, with 3-1, 2-2, and 1-3 red-blue polarities. Out of the 25 events in Figure \ref{fig:alaska_sequences}, 22 are present in the USGS ComCat, while 3 were newly detected, for which their magnitudes were estimated using amplitude ratios with respect to the reference event in the sequence.

\subsection{Vrancea}

The Vrancea seismicity nest under the Romanian Carpathians is one of the three major intermediate depth seismicity nests recognized by \cite{ZARIFI2003237,Prieto2012}. The Vrancea region has seen several Mw 7+ events in the last century, including the 1940 Mw 7.7, and the 1977 Mw 7.5 earthquakes, the most destructive in the country's history \citep{FUCHS1979225}. The Vrancea nest, has an extent of 70 km by 30 km and is elongated in the northeast to southwest direction. It has earthquakes down to depths of nearly 200 km that are disconnected from the shallow seismicity. This location is known to have a heterogeneous stress state based on the distribution of focal mechanisms \citep{PETRESCU2021228688,Craiu2022}, as a consequence of a relic slab from subduction that ceased 8 Ma ago, which is nowadays near vertical \citep{MATENCO2016807}. The stress state and faulting style have been linked to the stability fields of serpentine \citep{Ferrand2017}.

We conducted template matching using the events reported by the Romanian National Institute for Earth and Physics (NIEP) as templates and scanned them on continuous data from the Romanian Seismic Network \citep{https://doi.org/10.7914/sn/ro} .  We found only one pair of anti-similar events events occurring in 2010 and 2019, at a depth of 148 km. The events in these sequence are a magnitude 3.7 event and a newly detected event. This pair of events is located within the Vrancea seismic nest, as indicated in Figure \ref{fig:world}. The reports for this location are much less numerous than Alaska and Colombia, which may be a consequence of the limited spatial footprint of the Vrancea seismic nest compared to the extent of the Wadatti-Benioff zones of both Alaska and Colombia. As more data is recorded and we can extend our analysis, more events should be uncovered. This is the third of the main three intermediate depth seismic nests, alongside the Bucaramanga Nest \citep{Prieto2012} and the Hindu Kush \citep{cesca2026} where anti-similar earthquakes have been reported. 

\begin{figure}
    \centering
    \includegraphics[width=0.95\linewidth]{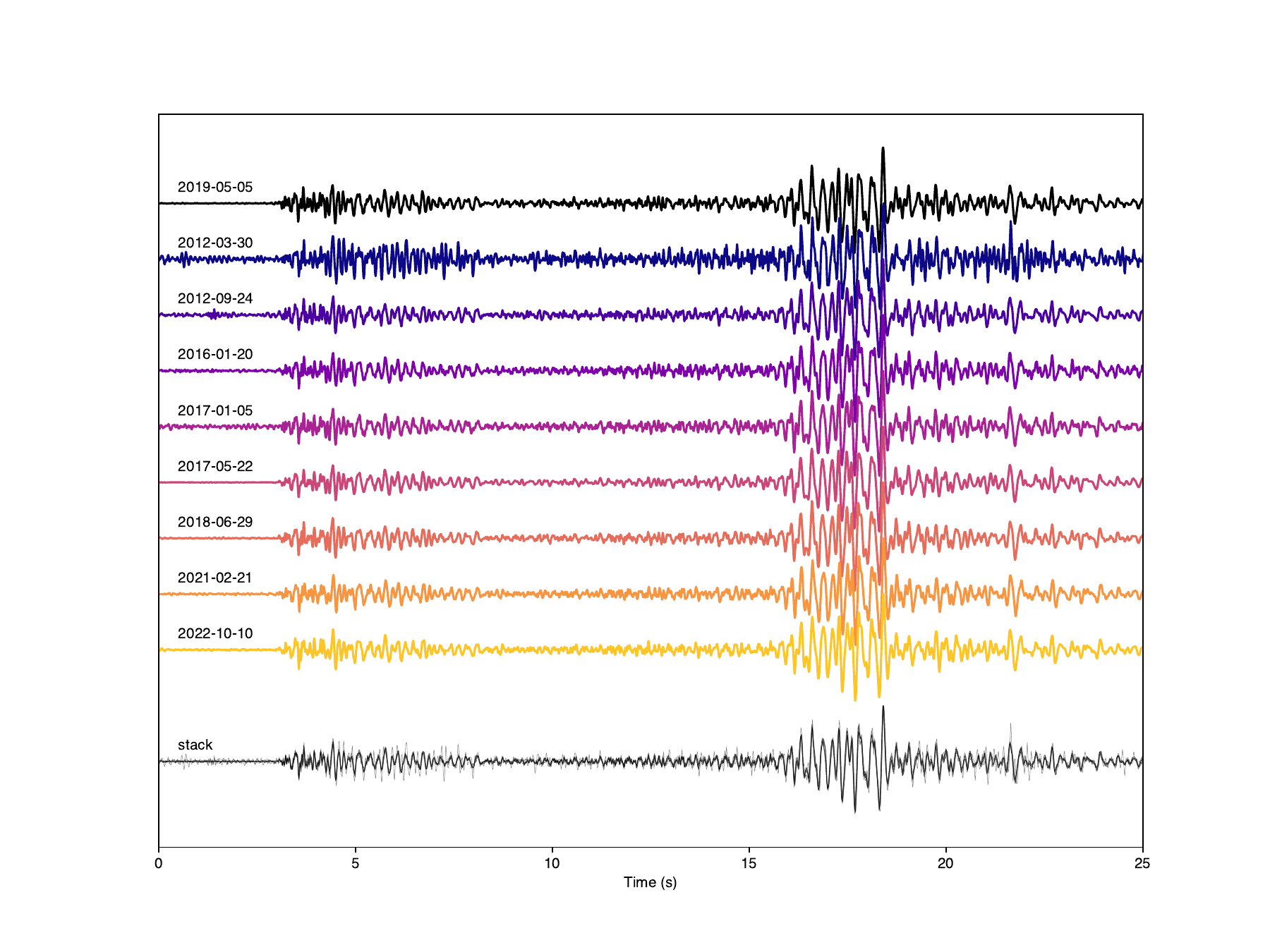}
    \caption{Highly correlated earthquakes in the Vrancea Nest as recorded by station RO.VRI.}
    \label{fig:vrancea}
\end{figure}

On the other hand, we identified several sequences with highly similar earthquakes, perhaps with some repeating earthquakes. The waveforms shown in Figure \ref{fig:vrancea} is a sequence that was active for over a decade and includes nine earthquakes between 2012 and 2022, with magnitudes ranging between 2.8 and 3.2. This is an analog observation to other subduction zones, where earthquakes in close proximity, but with no overlapping ruptures generate highly similar waveforms \citep{TSUCHIYAMA2021106695}.

\section{Discussion}

We documented anti-similar earthquake sequences at intermediate depth in Colombia, Alaska, and Romania (Figure \ref{fig:world}), all related to active or relict subduction, which combined with other reports from Japan \citep{nakajima2013}, Alaska \citep{plourde_2019}, Chile \citep{folesky_anti}, Hindu Kush \citep{cesca2026}, indicate that anti-similar earthquakes are a globally prevalent phenomenon.

In Colombia, the majority of these sequences take place in the Bucaramanga Nest and its vicinity, some containing events above magnitude 4, and displaying a variety of durations, ranging from weeks to years. These sequences also show a variety of focal mechanisms (Figure \ref{fig:bmanga}). The depth distribution of anti-similar sequences is concentrated in the 140 to 150 km, with a few sequences as shallow as 100 km. We report a total of 5,985 earthquakes distributed in 992 sequences, a significant expansion to those previously reported by \cite{Prieto2012,Barrett2015}. Overall, 7 percent of the total seismicity in the Bucaramanga nest was found to belong to an anti-similar sequence.  \citep{Prieto2012} attributed the sequences in their study to subparallel faults within the nest, while  \cite{Barrett2015} found that one of these sequences occurs on conjugate fault planes in zones of boudinage within the nest, although both studies quantify their relocation uncertainties to be about the same as the separation between the planes.

The second most productive location is the Cucunubá cluster, with hundreds of earthquakes in anti-similar sequences, which is a recently recognized host of this type of earthquakes \citep{2025AGUFMS43D.0296A,prieto2026}. This cluster is at a depth of 140-150 km, similar to that of the Bucaramanga Nest, and sits at the southern and deep end of the Bucaramanga WBZ. The Bucaramanga nest and the Cucunubá cluster account for all the sequences found in the Bucaramanga WBZ. On the Cauca WBZ, we found sequences in the Cauca cluster \citep{chang2017}, at shallower depths than the other locations, around 100 km depth, which might be a consequence of a nearly slab perpendicular cluster of seismicity that has been linked to hydrofracture induced by fluids migrating out of the Nazca slab \citep{bishop2025}. This hydrofracture mechanism is another explanation for anti-similar sequences in other settings \citep{Cesca2024}. This cluster is also known to have a highly heterogeneous set of focal mechanisms \citep{Chang2019}. There are two more groups of anti-similar sequences farther south in the Cauca WBZ, that in common fashion to Bucaramanga and Cucunubá sit closer to the termination of slab seismicity at nearly 150 km depth.

In Alaska, we report 6 anti-similar earthquake sequences, with 4 of these sequences organized in two pairs that are only a few km apart. These sequences are active for the 5 year duration of our analysis, with the largest event in one of these sequences being $M_{L}$ 2.3 (Figure \ref{fig:alaska_sequences}). These sequences locate near the deep termination of the slab seismicity, and the majority on the flat slab section in southeast Alaska (Figure \ref{fig:world}). In total we report 30 events, with our easternmost group being very close to the report of \cite{plourde_2019}, and representing smaller magnitudes.

The search for anti-similar earthquakes in the Vrancea nest in Romania yielded only a single pair of  earthquakes at 148 km depth, similar to those in Alaska and Colombia. The spatial extent of the Vrancea nest may be limiting the number of occurrences, at least when compared with the hundreds of km of along-strike continuous WBZ in Alaska and Colombia.  Regardless of the low number, this report is significant because it represents the third out of the main three intermediate depth seismicity nests \citep{ZARIFI2003237} that hosts this type of earthquakes, considering that anti-similar sequences are known in the Bucaramanga nest \citep{Prieto2012}, and the Hindu Kush nest \citep{cesca2026}.

\cite{nakajima2013} analyzed 106 earthquakes of which 30 were found to correspond to 4 anti-similar sequences, with magnitudes in the range 2 to 3.5 at depths of 155 km. \cite{plourde_2019} analyzed 16 events, with 7 anti-similar events distributed in two sequences, one quartet and one triplet, each with only one anti-similar occurrence, with magnitudes in the range 3 to 5.4 at depths of 125 km. \cite{folesky_anti}  reported 4 anti-similar sequences, three pairs and one sequence of 38 events including 5 reverse polarity occurrences, with magnitudes in the range 2 to 3.6 at depths of 133 km. These reports combined account for 81 events. We report 6,767 events in anti-similar sequences, contributing  two orders of magnitude more observations of anti-similar events, alas dominated by the Bucaramanga Nest.

For Colombia and Alaska, around 10 percent of the events associated with anti-similar sequences are events newly detected by our template matching workflow, which is not a significant contribution in the number of events. The majority of the events that met the conditions to be assigned to an anti-similar sequences, that is cross correlation over multiple stations less than -0.8 or more than 0.8 might bias our analysis to larger earthquakes that are well recorded in multiple stations, and might prevent small events that have low amplitude and noisy waveforms from being assigned to anti-similar sequences.

The majority of these anti-similar sequences are located near the deep termination of slab seismicity, and in regions of high earthquake concentration, which is a commonality for all intermediate depth reports \citep{Prieto2012,nakajima2013,plourde_2019,folesky_anti}. These events span the magnitude range from 1 to 5, display a variety of temporal behaviors  (Figures \ref{fig:bmanga},\ref{fig:alaska_sequences}), and exhibit variability in focal mechanisms (Figure \ref{fig:bmanga}). Furthermore, the proportions between one polarity and the opposite are also widely varied, including sequences that are dominated by one polarity, and others that have subequal numbers of opposite polarity members Figures \ref{fig:bmanga},\ref{fig:alaska_sequences}). The consensus among these studies is that events with reverse polarities take place on separate faults, likely parallel, subparallel, or conjugate fault planes. Moreover, they may be related to metamorphic reactions in downgoing slabs, such as eclogitization of gabbro \citep{nakajima2013, plourde_2019}. The widespread occurrence of these anti-similar earthquakes and their link to metamorphic reactions may indicate a potential to track both the extent and rate of such reactions in subduction zones.

The studies of \cite{folesky_anti} and ours are the only large scale similarity searches that reported anti-similar earthquakes. On the other hand, the studies of \cite{nakajima2013} and \cite{plourde_2019} are focused analyses of one single cluster, while also only considering M2+ events. However numerous our report is, it is limited by the duration and availability of continuous data. In the case of Colombia and the Bucaramanga Nest, changes in instrumentation, especially in key stations BAR2 and BRR compromise the temporal consistency and detectability of anti-similar events. The same applies to Alaska and Romania. Future searches for more of these events in similar tectonic settings should reveal a more complete picture of the geographic distribution of anti-similar earthquakes at intermediate depth. \cite{nakajima2013} point out similar nearby intermediate depth seismicity clusters north and east of their study region. \cite{folesky_anti} proved a fraction of the subduction zone, and there are still thousands of km of slab seismicity to analyze. Other regions with abundant intermediate depth seismicity might reveal more of these sequences.

Short scale stress field variations have been invoked to explain the seemingly opposite direction of faulting. Such stress field orientations have been documented in volcanic settings \citep{Snadanbata_2022,Olive2024} and also in the vicinity of the San Andreas fault \citep{shamir_1988}. However, there are simpler geometrical arguments that explain the observations, such as slight differences in fault orientation \citep{zobackberoza93}, nearly orthogonal faults \citep{Trugman2020}, or subparallel faults \citep{Lu2025,lu2026}.

We also note that agreeing on a unifying language is  important for recognizing reports of anti-similar earthquakes. The study of \cite{nakajima2013} did not use the term anti-repeating or anti-similar or flipped polarity and was thus nor recognized in the review of \cite{Cesca2024}. The same applies to the study of \cite{plourde_2019} that uses the term anti-parallel earthquakes, which is also not recognized in the review of \cite{Cesca2024}.

Our reports on several locations with abundant intermediate depth seismicity in Colombia beyond those by \cite{Prieto2012,prieto2026}, several locations in Alaska beyond those by \cite{plourde_2019}, combined with those in Vrancea, complemented by the reports of \cite{nakajima2013} in Japan, \cite{folesky_anti} in Chile, plus those in \cite{cesca2026} indicate a globally prevalent process at subduction zones, that is likely linked to metamorphic reactions in downgoing slabs. The locations, rates, and magnitudes of these earthquake sequences could track the rate of those metamorphic reactions.

\vspace{10mm}
\section{Conclusions}

Enhanced earthquake catalogs using a combination of machine learning and template matching enabled the discovery of anti-similar earthquakes at intermediate depth in Colombia, Alaska, and Romania. These observations, and similar observations from Japan and Chile, indicate that anti-similar earthquakes are widespread in subduction zones at intermediate depth. The temporal recurrence patterns include both short- to long-lived sequences, with the number of anti-similar events also ranging from one event to half the events within a sequence. Future searches for such events in other regions with intense intermediate depth seismicity should help understand their global prevalence, significance, and implications for the underlying mechanism(s). 



\section*{Data and Resources}
We used continuous data from:
Servicio Geológico Colombiano 
\citep{https://doi.org/10.7914/sn/cm}
Alaska Geophysical Network \citep{https://doi.org/10.7914/sn/ak}, Alaska Volcano Observatory \citep{av},
Romanian Seismic Network
\citep{https://doi.org/10.7914/sn/ro}
and their corresponding earthquake catalogs. The SGC earthquake catalog was consulted here \url{https://www.sgc.gov.co/catalogo} and the focal mechanisms here \url{https://bdrsnc.sgc.gov.co/sismologia1/sismologia/focal_seiscomp_3/index.html}

\section*{Declaration of Competing Interests}
The authors declare no conflicts of interest. This is an LLM free manuscript.


\newpage
\bibliographystyle{apalike}
\bibliography{antisimilar}

@article{shamir_1988,
author = {Shamir, Gadi and Zoback, Mark D. and Barton, Colleen A.},
title = {In situ stress orientation near the San Andreas Fault: Preliminary results to 2.1 km depth from the Cajon Pass Scientific Drillhole},
journal = {Geophysical Research Letters},
volume = {15},
number = {9},
pages = {989-992},
doi = {https://doi.org/10.1029/GL015i009p00989},
url = {https://agupubs.onlinelibrary.wiley.com/doi/abs/10.1029/GL015i009p00989},
eprint = {https://agupubs.onlinelibrary.wiley.com/doi/pdf/10.1029/GL015i009p00989},
year = {1988}
}

@INPROCEEDINGS{2025AGUFMS43D.0296A,
       author = {{Aguilar Suarez}, Albert and {Beroza}, Gregory C. and {Monsalve}, Gaspar and {Pedraza}, Patricia and {Prieto}, German A. and {Wagner}, Lara S.},
        title = "{Pervasive Anti-repeating Earthquakes at Intermediate Depth in Colombia}",
    booktitle = {AGU Fall Meeting Abstracts},
         year = 2025,
       series = {AGU Fall Meeting Abstracts},
       volume = {2025},
        month = dec,
          eid = {S43D-0296},
        pages = {S43D-0296},
       adsurl = {https://ui.adsabs.harvard.edu/abs/2025AGUFMS43D.0296A}
}

@article{ye2014,
author = {Ye, Lingling and Lay, Thorne and Kanamori, Hiroo},
title = {The 23 June 2014 Mw 7.9 Rat Islands archipelago, Alaska, intermediate depth earthquake},
journal = {Geophysical Research Letters},
volume = {41},
number = {18},
pages = {6389-6395},
doi = {https://doi.org/10.1002/2014GL061153},
url = {https://agupubs.onlinelibrary.wiley.com/doi/abs/10.1002/2014GL061153},
eprint = {https://agupubs.onlinelibrary.wiley.com/doi/pdf/10.1002/2014GL061153},
year = {2014}
}

@misc{av,
  doi = {10.7914/SN/AV},
  url = {https://www.fdsn.org/networks/detail/AV/},
  author = {{Alaska Volcano Observatory/USGS}},
  title = {Alaska Volcano Observatory},
  publisher = {International Federation of Digital Seismograph Networks},
  year = {1988}
}

@article{bauer2014,
    author = {Bauer, Mark A. and Pavlis, Gary L. and Landes, Michael},
    title = {Subduction geometry of the Yakutat terrane, southeastern Alaska},
    journal = {Geosphere},
    volume = {10},
    number = {6},
    pages = {1161-1176},
    year = {2014},
    month = {12},
    issn = {1553-040X},
    doi = {10.1130/GES00852.1},
    url = {https://doi.org/10.1130/GES00852.1},
    eprint = {https://pubs.geoscienceworld.org/gsa/geosphere/article-pdf/10/6/1161/3335832/1161.pdf},
}

@article{bishop2025,
author = {Bishop, Brandon T. and Warren, Linda M. and Aravena, Pablo and Cho, Sungwon and Soto-Cordero, Lillian and Pedraza, Patricia and Prieto, Germán A. and Dionicio, Viviana},
title = {The Deep Lithospheric Structure of Terrane Accretion as Revealed Through Patterns of Seismicity Associated With the Collision of the Panamá–Chocó Block and South America Beneath Cauca, Colombia},
journal = {Journal of Geophysical Research: Solid Earth},
volume = {130},
number = {6},
pages = {e2024JB030067},
doi = {https://doi.org/10.1029/2024JB030067},
url = {https://agupubs.onlinelibrary.wiley.com/doi/abs/10.1029/2024JB030067},
eprint = {https://agupubs.onlinelibrary.wiley.com/doi/pdf/10.1029/2024JB030067},
note = {e2024JB030067 2024JB030067},
year = {2025}
}

@incollection{ZARIFI2003237,
title = {CHARACTERISTICS OF DENSE NESTS OF DEEP AND INTERMEDIATE-DEPTH SEISMICITY},
series = {Advances in Geophysics},
publisher = {Elsevier},
volume = {46},
pages = {237-278},
year = {2003},
issn = {0065-2687},
doi = {https://doi.org/10.1016/S0065-2687(03)46004-4},
url = {https://www.sciencedirect.com/science/article/pii/S0065268703460044},
author = {Zoya Zarifi and Jens Havskov}
}

@article{shuler_2013,
author = {Shuler, Ashley and Ekström, Göran and Nettles, Meredith},
title = {Physical mechanisms for vertical-CLVD earthquakes at active volcanoes},
journal = {Journal of Geophysical Research: Solid Earth},
volume = {118},
number = {4},
pages = {1569-1586},
doi = {https://doi.org/10.1002/jgrb.50131},
url = {https://agupubs.onlinelibrary.wiley.com/doi/abs/10.1002/jgrb.50131},
eprint = {https://agupubs.onlinelibrary.wiley.com/doi/pdf/10.1002/jgrb.50131},
year = {2013}
}

@article{Snadanbata_2022,
author = {Sandanbata, Osamu and Watada, Shingo and Satake, Kenji and Kanamori, Hiroo and Rivera, Luis and Zhan, Zhongwen},
title = {Sub-Decadal Volcanic Tsunamis Due To Submarine Trapdoor Faulting at Sumisu Caldera in the Izu–Bonin Arc},
journal = {Journal of Geophysical Research: Solid Earth},
volume = {127},
number = {9},
pages = {e2022JB024213},
doi = {https://doi.org/10.1029/2022JB024213},
url = {https://agupubs.onlinelibrary.wiley.com/doi/abs/10.1029/2022JB024213},
eprint = {https://agupubs.onlinelibrary.wiley.com/doi/pdf/10.1029/2022JB024213},
note = {e2022JB024213 2022JB024213},
year = {2022}
}

@Article{Cesca2020,
author={Cesca, Simone
and Letort, Jean
and Razafindrakoto, Hoby N. T.
and Heimann, Sebastian
and Rivalta, Eleonora
and Isken, Marius P.
and Nikkhoo, Mehdi
and Passarelli, Luigi
and Petersen, Gesa M.
and Cotton, Fabrice
and Dahm, Torsten},
title={Drainage of a deep magma reservoir near Mayotte inferred from seismicity and deformation},
journal={Nature Geoscience},
year={2020},
month={Jan},
day={01},
volume={13},
number={1},
pages={87-93},
issn={1752-0908},
doi={10.1038/s41561-019-0505-5},
url={https://doi.org/10.1038/s41561-019-0505-5}
}

@article{FUCHS1979225,
title = {The Romanian earthquake of March 4, 1977 ii. Aftershocks and migration of seismic activity +},
journal = {Tectonophysics},
volume = {53},
number = {3},
pages = {225-247},
year = {1979},
note = {Proceedings of the 16th General Assemble of the European Seismological Commission},
issn = {0040-1951},
doi = {https://doi.org/10.1016/0040-1951(79)90068-4},
url = {https://www.sciencedirect.com/science/article/pii/0040195179900684},
author = {K. Fuchs and K.-P. Bonjer and G. Bock and I. Cornea and C. Radu and D. Enescu and D. Jianu and A. Nourescu and G. Merkler and T. Moldoveanu and G. Tudorache}
}

@article{plourde_2019,
    author = {Plourde, Alexandre P and Bostock, Michael G},
    title = {Relative moment tensors and deep Yakutat seismicity},
    journal = {Geophysical Journal International},
    volume = {219},
    number = {2},
    pages = {1447-1462},
    year = {2019},
    month = {11},
    issn = {0956-540X},
    doi = {10.1093/gji/ggz375},
    url = {https://doi.org/10.1093/gji/ggz375},
    eprint = {https://academic.oup.com/gji/article-pdf/219/2/1447/29807103/ggz375_supplemental_file.pdf},
}

@article{Jung2004,
author={Jung, Haemyeong
and Green II, Harry W.
and Dobrzhinetskaya, Larissa F.},
title={Intermediate-depth earthquake faulting by dehydration embrittlement with negative volume change},
journal={Nature},
year={2004},
month={Apr},
day={01},
volume={428},
number={6982},
pages={545-549},
issn={1476-4687},
doi={10.1038/nature02412},
url={https://doi.org/10.1038/nature02412}
}

@article{wei2020_alaska,
title = {Along-strike variations in intermediate-depth seismicity and arc magmatism along the Alaska Peninsula},
journal = {Earth and Planetary Science Letters},
volume = {563},
pages = {116878},
year = {2021},
issn = {0012-821X},
doi = {https://doi.org/10.1016/j.epsl.2021.116878},
url = {https://www.sciencedirect.com/science/article/pii/S0012821X21001370},
author = {S. Shawn Wei and Philipp Ruprecht and Sydney L. Gable and Ellyn G. Huggins and Natalia Ruppert and Lei Gao and Haijiang Zhang}
}

@article{zanjani2022,
    author = {Aziz Zanjani, Farzaneh and Lin, Guoqing},
    title = {Double Seismic Zones along the Eastern Aleutian‐Alaska Subduction Zone Revealed by a High‐Precision Earthquake Relocation Catalog},
    journal = {Seismological Research Letters},
    volume = {93},
    number = {5},
    pages = {2753-2769},
    year = {2022},
    month = {07},
    issn = {0895-0695},
    doi = {10.1785/0220210348},
    url = {https://doi.org/10.1785/0220210348},
    eprint = {https://pubs.geoscienceworld.org/ssa/srl/article-pdf/93/5/2753/5681999/srl-2021348.1.pdf},
}

@inbook{ruppert_2024,
author = {Ruppert, Natalia A.},
publisher = {American Geophysical Union (AGU)},
isbn = {9781394195947},
title = {A Decade of Alaska Seismicity: 2013–2022},
booktitle = {Tectonics and Seismic Structure of Alaska and Northwestern Canada},
chapter = {3},
pages = {45-84},
doi = {https://doi.org/10.1002/9781394195947.ch3},
url = {https://agupubs.onlinelibrary.wiley.com/doi/abs/10.1002/9781394195947.ch3},
eprint = {https://agupubs.onlinelibrary.wiley.com/doi/pdf/10.1002/9781394195947.ch3},
year = {2024}
}

@article{MATENCO2016807,
title = {The interplay between tectonics, sediment dynamics and gateways evolution in the Danube system from the Pannonian Basin to the western Black Sea},
journal = {Science of The Total Environment},
volume = {543},
pages = {807-827},
year = {2016},
issn = {0048-9697},
doi = {https://doi.org/10.1016/j.scitotenv.2015.10.081},
url = {https://www.sciencedirect.com/science/article/pii/S0048969715308974},
author = {Liviu Matenco and Ioan Munteanu and Marten {ter Borgh} and Adrian Stanica and Marius Tilita and Gilles Lericolais and Corneliu Dinu and Gheorghe Oaie}
}

@article{Perez-Forero2023,
author={Pérez-Forero, Diego
and Koulakov, Ivan
and Vargas, Carlos A.
and Gerya, Taras
and Al Arifi, Nassir},
title={Lithospheric delamination as the driving mechanism of intermediate-depth seismicity in the Bucaramanga Nest, Colombia},
journal={Scientific Reports},
year={2023},
month={Dec},
day={27},
volume={13},
number={1},
pages={23084},
issn={2045-2322},
doi={10.1038/s41598-023-50159-4},
url={https://doi.org/10.1038/s41598-023-50159-4}
}

@book{ramirez1974,
  title     = "Historia de los terremotos en Colombia",
  author    = "Ramírez, Jesús Emilio",
  year      = 1974,
  publisher = "República de Colombia, Instituto Geográfico Agustin Codazzi, Subdirección de Investigaciones y Divulgación Geográfica",
  address   = "Bogotá"
}

@article{dionicio_fm,
    author = {Dionicio, Viviana and Pedraza, Patricia and Poveda, Esteban},
    title = {Moment tensor and focal mechanism data of earthquakes recorded by Servicio Geológico Colombiano from 2014 to 2021. },
    journal = {Boletín Geológico},
    year = {2023}
}

@article{Frohlich1995,
author = {Frohlich, Cliff and Kadinsky-Cade, Katharine and Davis, Scott D.},
isbn = {0875532071},
journal = {Bulletin of the Seismological Society of America},
number = {6},
pages = {1622--1634},
title = {{A Reexamination of the Bucaramanga, Colombiam Earthquake Nest}},
volume = {85},
year = {1995}
}

@article{BECK1998115,
title = {Source characteristics of historic earthquakes along the central Chile subduction Askew et alzone},
journal = {Journal of South American Earth Sciences},
volume = {11},
number = {2},
pages = {115-129},
year = {1998},
issn = {0895-9811},
doi = {https://doi.org/10.1016/S0895-9811(98)00005-4},
url = {https://www.sciencedirect.com/science/article/pii/S0895981198000054},
author = {S. Beck and S. Barrientos and E. Kausel and M. Reyes}
}

@article{Zhan2020,
author = {Zhan, Zhongwen},
doi = {10.1146/annurev-earth-053018-060314},
issn = {00846597},
journal = {Annual Review of Earth and Planetary Sciences},
pages = {147--174},
title = {{Mechanisms and Implications of Deep Earthquakes}},
volume = {48},
year = {2020}
}

@article{Green1995,
author = {Green, Harry W. and Houston, Heidi},
doi = {10.1146/annurev.ea.23.050195.001125},
issn = {0084-6597},
journal = {Annual Review of Earth and Planetary Sciences},
month = {may},
number = {1},
pages = {169--213},
title = {{The Mechanics of Deep Earthquakes}},
url = {http://www.annualreviews.org/doi/10.1146/annurev.ea.23.050195.001125},
volume = {23},
year = {1995}
}

@phdthesis{Barrett2015,
author = {Barrett, Sarah Anne},
number = {August},
school = {Stanford University},
title = {{Seismological Constraints on the Mechanics of Intermediate Depth Earthquakes in the Bucaramanga Nest}},
url = {http://purl.stanford.edu/bn324ff7907},
year = {2015}
}

@article{wagnerpanama_2025,
author = {Wagner, Lara S. and Prieto, German A. and Montes, Camilo and Ramos, J. P. and Dionicio, V. and Pedraza, P.},
title = {Breaking the Caribbean Plate: Subduction Initiation Beneath the Northern Margin of Panama},
journal = {Geophysical Research Letters},
volume = {52},
number = {18},
pages = {e2025GL116734},
doi = {https://doi.org/10.1029/2025GL116734},
url = {https://agupubs.onlinelibrary.wiley.com/doi/abs/10.1029/2025GL116734},
eprint = {https://agupubs.onlinelibrary.wiley.com/doi/pdf/10.1029/2025GL116734},
note = {e2025GL116734 2025GL116734},
year = {2025}
}

@article{Ferrand2017,
author = {Ferrand, Thomas P. and Hilairet, Nad{\`{e}}ge and Incel, Sarah and Deldicque, Damien and Labrousse, Lo{\"{i}}c and Gasc, Julien and Renner, Joerg and Wang, Yanbin and Green, Harry W. and Schubnel, Alexandre},
doi = {10.1038/ncomms15247},
issn = {20411723},
journal = {Nature Communications},
number = {May},
pages = {1--11},
pmid = {28504263},
title = {{Dehydration-driven stress transfer triggers intermediate-depth earthquakes}},
volume = {8},
year = {2017}
}

@Article{Craiu2022,
author={Craiu, Andreea
and Ferrand, Thomas P.
and Manea, Elena F.
and Vrijmoed, Johannes C.
and M{\u{a}}rmureanu, Alexandru},
title={A switch from horizontal compression to vertical extension in the Vrancea slab explained by the volume reduction of serpentine dehydration},
journal={Scientific Reports},
year={2022},
month={Dec},
day={24},
volume={12},
number={1},
pages={22320},
issn={2045-2322},
doi={10.1038/s41598-022-26260-5},
url={https://doi.org/10.1038/s41598-022-26260-5}
}

@article{PETRESCU2021228688,
title = {Tectonic regimes and stress patterns in the Vrancea Seismic Zone: Insights into intermediate-depth earthquake nests in locked collisional settings},
journal = {Tectonophysics},
volume = {799},
pages = {228688},
year = {2021},
issn = {0040-1951},
doi = {https://doi.org/10.1016/j.tecto.2020.228688},
url = {https://www.sciencedirect.com/science/article/pii/S0040195120303711},
author = {Laura Petrescu and Felix Borleanu and Mircea Radulian and Alik Ismail-Zadeh and Liviu Maţenco}
}

@article{Poli2016,
author = {Poli, P. and Prieto, G. A. and Yu, C. Q. and Florez, M. and Agurto-Detzel, H. and Mikesell, T. D. and Chen, G. and Dionicio, V. and Pedraza, P.},
doi = {10.1093/gji/ggw065},
issn = {1365246X},
journal = {Geophysical Journal International},
number = {2},
pages = {988--994},
title = {{Complex rupture of the M6.3 2015 March 10 Bucaramanga earthquake: Evidence of strong weakening process}},
volume = {205},
year = {2016}
}

@article{TSUCHIYAMA2021106695,
title = {Diversity of deep earthquakes with waveform similarity},
journal = {Physics of the Earth and Planetary Interiors},
volume = {314},
pages = {106695},
year = {2021},
issn = {0031-9201},
doi = {https://doi.org/10.1016/j.pepi.2021.106695},
url = {https://www.sciencedirect.com/science/article/pii/S0031920121000534},
author = {Ayako Tsuchiyama and Junichi Nakajima}
}

@incollection{Vargas2020,
author = {Vargas, Carlos A.},
booktitle = {Subduction geometries},
number = {May},
pages = {306},
title = {{Chapter 11 Subduction geometries}},
volume = {4},
year = {2020}
}

@article{chang2017,
    author = {Chang, Ying and Warren, Linda M. and Prieto, Germán A.},
    title = {Precise Locations for Intermediate‐Depth Earthquakes in the Cauca Cluster, Colombia},
    journal = {Bulletin of the Seismological Society of America},
    volume = {107},
    number = {6},
    pages = {2649-2663},
    year = {2017},
    month = {10},
    issn = {0037-1106},
    doi = {10.1785/0120170127},
    url = {https://doi.org/10.1785/0120170127},
    eprint = {https://pubs.geoscienceworld.org/ssa/bssa/article-pdf/107/6/2649/3992794/bssa-2017127.1.pdf},
}

@article{Chang2019,
author = {Chang, Y. and Warren, L. M. and Zhu, L. and Prieto, G. A.},
doi = {10.1029/2018JB016804},
issn = {21699356},
journal = {Journal of Geophysical Research: Solid Earth},
number = {1},
pages = {822--836},
title = {{Earthquake Focal Mechanisms and Stress Field for the Intermediate-Depth Cauca Cluster, Colombia}},
volume = {124},
year = {2019}
}

@article{Wagner2017,
author = {Wagner, L. S. and Jaramillo, J. S. and Ram{\'{i}}rez-Hoyos, L. F. and Monsalve, G. and Cardona, A. and Becker, T. W.},
doi = {10.1002/2017GL073981},
issn = {19448007},
journal = {Geophysical Research Letters},
month = {jul},
number = {13},
pages = {6616--6623},
publisher = {Blackwell Publishing Ltd},
title = {{Transient slab flattening beneath Colombia}},
volume = {44},
year = {2017}
}

@article{Vargas2013,
author = {Vargas, Carlos A. and Mann, Paul},
doi = {10.1785/0120120328},
issn = {00371106},
journal = {Bulletin of the Seismological Society of America},
month = {jun},
number = {3},
pages = {2025--2046},
title = {{Tearing and breaking off of subducted slabs as the result of collision of the panama arc-indenter with Northwestern South America}},
volume = {103},
year = {2013}
}

@article{zobackberoza93,
    author = {Zoback, Mark D. and Beroza, Gregory C.},
    title = {Evidence for near-frictionless faulting in the 1989 (M 6.9) Loma Prieta, California, earthquake and its aftershocks},
    journal = {Geology},
    volume = {21},
    number = {2},
    pages = {181-185},
    year = {1993},
    month = {02},
    issn = {0091-7613},
    doi = {10.1130/0091-7613(1993)021<0181:EFNFFI>2.3.CO;2},
    url = {https://doi.org/10.1130/0091-7613(1993)021<0181:EFNFFI>2.3.CO;2},
    eprint = {https://pubs.geoscienceworld.org/gsa/geology/article-pdf/21/2/181/3514699/i0091-7613-21-2-181.pdf},
}

@article{Prieto2012,
author = {Prieto, Germ{\'{a}}n A. and Beroza, Gregory C. and Barrett, Sarah A. and L{\'{o}}pez, Gabriel A. and Florez, Manuel},
doi = {10.1016/j.tecto.2012.07.019},
issn = {00401951},
journal = {Tectonophysics},
pages = {42--56},
publisher = {Elsevier B.V.},
title = {{Earthquake nests as natural laboratories for the study of intermediate-depth earthquake mechanics}},
url = {http://dx.doi.org/10.1016/j.tecto.2012.07.019},
volume = {570-571},
year = {2012}
}

@article{nakajima2013,
    author = {Nakajima, Junichi and Uchida, Naoki and Shiina, Takahiro and Hasegawa, Akira and Hacker, Bradley R. and Kirby, Stephen H.},
    title = {Intermediate-depth earthquakes facilitated by eclogitization-related stresses},
    journal = {Geology},
    volume = {41},
    number = {6},
    pages = {659-662},
    year = {2013},
    month = {06},
    issn = {0091-7613},
    doi = {10.1130/G33796.1},
    url = {https://doi.org/10.1130/G33796.1},
    eprint = {https://pubs.geoscienceworld.org/gsa/geology/article-pdf/41/6/659/3544654/659.pdf},
}

@article{folesky_anti,
    author = {Folesky, Jonas and Kummerow, Jörn and Petersen, Gesa and Jan Hoffman, Laurens},
    title = {Reverse‐Polarity Earthquakes at Intermediate Depth in North Chile and Bolivia: Observations of a Rare Phenomenon},
    journal = {The Seismic Record},
    volume = {6},
    number = {3},
    pages = {329-337},
    year = {2026},
    month = {07},
    issn = {2694-4006},
    doi = {10.1785/0320260015},
    url = {https://doi.org/10.1785/0320260015},
    eprint = {https://pubs.geoscienceworld.org/ssa/tsr/article-pdf/6/3/329/8123395/tsr-2026015.1.pdf},
}

@misc{https://doi.org/10.7914/sn/ak,
  doi = {10.7914/SN/AK},
  url = {https://www.fdsn.org/networks/detail/AK/},
  author = {{Alaska Earthquake Center, Univ. of Alaska Fairbanks}},
  title = {Alaska Geophysical Network},
  publisher = {International Federation of Digital Seismograph Networks},
  year = {1987}
}

@misc{https://doi.org/10.7914/sn/ro,
  doi = {10.7914/SN/RO},
  url = {https://www.fdsn.org/networks/detail/RO/},
  author = {{National Institute for Earth Physics}},
  title = {Romanian Seismic Network},
  publisher = {International Federation of Digital Seismograph Networks},
  year = {1994}
}

@misc{https://doi.org/10.7914/sn/cm,
  doi = {10.7914/SN/CM},
  url = {https://www.fdsn.org/networks/detail/CM/},
  author = {{Servicio Geológico Colombiano}},
  title = {Red Sismologica Nacional de Colombia},
  publisher = {International Federation of Digital Seismograph Networks},
  year = {1993}
}

@article{Cesca2024,
author = {Cesca, Simone and Niemz, Peter and Dahm, Torsten and Ide, Satoshi},
doi = {10.1038/s43247-024-01290-1},
isbn = {4324702401290},
issn = {26624435},
journal = {Communications Earth and Environment},
number = {1},
pages = {1--11},
publisher = {Springer US},
title = {{Anti-repeating earthquakes and how to explain them}},
volume = {5},
year = {2024}
}

@article{Shelly2016,
author = {Shelly, David R. and Hardebeck, Jeanne L. and Ellsworth, William L. and Hill, David P.},
doi = {10.1002/2016JB013437},
issn = {21699356},
journal = {Journal of Geophysical Research: Solid Earth},
number = {12},
pages = {8622--8641},
title = {{A new strategy for earthquake focal mechanisms using waveform-correlation-derived relative polarities and cluster analysis: Application to the 2014 Long Valley Caldera earthquake swarm}},
volume = {121},
year = {2016}
}

@article{Trugman2020,
author = {Trugman, Daniel T. and Ross, Zachary E. and Johnson, Paul A.},
doi = {10.1029/2019GL085888},
issn = {19448007},
journal = {Geophysical Research Letters},
number = {1},
pages = {1--8},
title = {{Imaging Stress and Faulting Complexity Through Earthquake Waveform Similarity}},
volume = {47},
year = {2020}
}

@article{Hauksson2005,
author = {Hauksson, Egill and Shearer, Peter},
doi = {10.1785/0120040167},
issn = {00371106},
journal = {Bulletin of the Seismological Society of America},
number = {3},
pages = {896--903},
title = {{Southern California hypocenter relocation with waveform cross-correlation, part 1: Results using the double-difference method}},
volume = {95},
year = {2005}
}

@article{ide2011,
author = {Satoshi Ide  and Annemarie Baltay  and Gregory C. Beroza },
title = {Shallow Dynamic Overshoot and Energetic Deep Rupture in the 2011 Mw 9.0 Tohoku-Oki Earthquake},
journal = {Science},
volume = {332},
number = {6036},
pages = {1426-1429},
year = {2011},
doi = {10.1126/science.1207020},
URL = {https://www.science.org/doi/abs/10.1126/science.1207020},
eprint = {https://www.science.org/doi/pdf/10.1126/science.1207020}}

@article{ AguilarSuarez_Beroza_2025,
title={Picking Regional Seismic Phase Arrival Times with Deep Learning}, volume={4}, url={https://seismica.library.mcgill.ca/article/view/1431}, DOI={10.26443/seismica.v4i1.1431}, abstractNote={&amp;lt;p&amp;gt;Sparse instrumental coverage for much of the Earth requires working with regional seismic phases for effective seismic monitoring. Machine learning phase pickers to date have focused on local earthquake recordings. Here we present deep learning models designed and trained to be effective at picking the arrival times of earthquake phases at distances up to 20 degrees. We trained our models on the CREW dataset, which includes 1.6 million earthquake waveforms with over 3.2 million labeled arrivals on 5 minute long three component seismograms. We present models that accurately pick the first arriving P and S waves and models that pick and classify Pn, Pg, Sn, and Sg phase arrivals. We apply these models in a variety of settings and compare their performance to established machine learning models that were trained on local earthquake recordings. We demonstrate the abilities of our models by finding new earthquakes in the Gorda plate offshore northern California. Finally, we use our multiple phase picker to find new examples with secondary arrivals from our massive training dataset. The goal of this method is to improve automatic earthquake monitoring in regions of sparse instrumental coverage and seismicity in remote regions far from instrumentation. &amp;lt;/p&amp;gt;}, number={1}, journal={Seismica}, author={Aguilar Suarez, Albert Leonardo and Beroza, Gregory}, year={2025}, month={Apr.} }

@article{eaton1970,
    author = {Eaton, J. P. and O'Neill, M. E. and Murdock, J. N.},
    title = {Aftershocks of the 1966 Parkfield-Cholame, California, earthquake: A detailed study*},
    journal = {Bulletin of the Seismological Society of America},
    volume = {60},
    number = {4},
    pages = {1151-1197},
    year = {1970},
    month = {08},
    issn = {0037-1106},
    doi = {10.1785/BSSA0600041151},
    url = {https://doi.org/10.1785/BSSA0600041151},
    eprint = {https://pubs.geoscienceworld.org/ssa/bssa/article-pdf/60/4/1151/5313123/bssa0600041151.pdf},
}

@article{Yagi_Fukahata_Okuwaki_Takagawa_Toda_2025, title={Breaking the Cycle: Short Recurrence and Overshoot of an M9-class Kamchatka Earthquake}, volume={4}, url={https://seismica.library.mcgill.ca/article/view/2012}, DOI={10.26443/seismica.v4i2.2012}, abstractNote={&amp;lt;p&amp;gt;M9-class megathrust earthquakes in subduction zones are generally thought to release slip deficits on the plate interface accumulated over centuries. However, the 2025 Kamchatka earthquake (Mw 8.8-8.9) ruptured nearly the same area as the 1952 Mw 9.0 event, as shown by the aftershock distribution. This unusually short recurrence interval challenges conventional seismic-cycle models. Using a cutting-edge source inversion technique, we analyze seismic data to estimate the spatiotemporal slip-rate evolution of the 2025 event. The results show that the 2025 rupture involved fault slips exceeding 9 m across a broad region from southern Kamchatka to the northern Kuril Islands, which is significantly greater than the plate convergence of about 6 m since 1952, matching the large-slip area of the 1952 event. Slip rates in the large-slip area accelerated twice, probably due to dynamic stress perturbations and complex frictional behaviour, and were followed by low-angle normal-faulting aftershocks suggesting dynamic overshoot. The results indicate that the 2025 earthquake released a substantial amount of the slip deficit that had not been released during the 1952 event. Therefore, the residual strains that remain after a great earthqauke and are not considered in current hazard forecasting can lead to shorter recurrence. This finding offers important clues to how great earthquakes release slip deficits and may help develop more physically based long-term forecasts.&amp;lt;/p&amp;gt;}, number={2}, journal={Seismica}, author={Yagi, Yuji and Fukahata, Yukitoshi and Okuwaki, Ryo and Takagawa, Tomohiro and Toda, Shinji}, year={2025}, month={Nov.} }

@Article{Lu2025,
author={Lu, Weifan
and Ide, Satoshi
and Yue, Han},
title={Small-scale stress heterogeneity inferred many anti-repeating earthquakes in  Sierra Valley, Nevada},
journal={Earth, Planets and Space},
year={2025},
month={Aug},
day={25},
volume={77},
number={1},
pages={142},
issn={1880-5981},
doi={10.1186/s40623-025-02273-y},
url={https://doi.org/10.1186/s40623-025-02273-y}
}

@article{cesca_2016,
    author = {Cesca, S. and Grigoli, F. and Heimann, S. and Dahm, T. and Kriegerowski, M. and Sobiesiak, M. and Tassara, C. and Olcay, M.},
    title = {The Mw 8.1 2014 Iquique, Chile, seismic sequence: a tale of foreshocks and aftershocks},
    journal = {Geophysical Journal International},
    volume = {204},
    number = {3},
    pages = {1766-1780},
    year = {2016},
    month = {02},
    issn = {0956-540X},
    doi = {10.1093/gji/ggv544},
    url = {https://doi.org/10.1093/gji/ggv544},
    eprint = {https://academic.oup.com/gji/article-pdf/204/3/1766/1983846/ggv544.pdf},
}

@article{lu2026,
author = {Lu, Weifan},
title = {Migration of Multimodal Deep Crustal Earthquake Swarm Beneath the Abu Volcano Group, Japan},
journal = {Geophysical Research Letters},
volume = {53},
number = {3},
pages = {e2025GL119316},
doi = {https://doi.org/10.1029/2025GL119316},
url = {https://agupubs.onlinelibrary.wiley.com/doi/abs/10.1029/2025GL119316},
eprint = {https://agupubs.onlinelibrary.wiley.com/doi/pdf/10.1029/2025GL119316},
note = {e2025GL119316 2025GL119316},
year = {2026}
}

@article{prieto2026,
    author = {Prieto, Germán A.},
    title = {Diversidad de los terremotos de profundidad intermedia en Colombia: Un laboratorio natural},
    journal = {Rev. Acad. Col. Ciencias Ex. Fis. Nat.},
    year = {2026}}

@article{cesca2026,
    author = {Cesca, Simone and Stich, Daniel and Ekström, Göran and Niemz, Peter and Moya Sánchez, Ricardo and Ide, Satoshi and Dahm, Torsten},
    title = {Opposite Faulting in Global Seismicity},
    journal = {The Seismic Record},
    volume = {6},
    number = {1},
    pages = {85-95},
    year = {2026},
    month = {03},
    issn = {2694-4006},
    doi = {10.1785/0320250052},
    url = {https://doi.org/10.1785/0320250052},
    eprint = {https://pubs.geoscienceworld.org/ssa/tsr/article-pdf/6/1/85/7784781/tsr-2025052.1.pdf},
}

@article{shearer2024,
author = {Shearer, Peter M. and Shabikay Senobari, Nader and Fialko, Yuri},
title = {Implications of a Reverse Polarity Earthquake Pair on Fault Friction and Stress Heterogeneity Near Ridgecrest, California},
journal = {Journal of Geophysical Research: Solid Earth},
volume = {129},
number = {11},
pages = {e2024JB029562},
doi = {https://doi.org/10.1029/2024JB029562},
url = {https://agupubs.onlinelibrary.wiley.com/doi/abs/10.1029/2024JB029562},
eprint = {https://agupubs.onlinelibrary.wiley.com/doi/pdf/10.1029/2024JB029562},
note = {e2024JB029562 2024JB029562},
year = {2024}
}

@article{eqcorrscan,
    author = {Chamberlain, Calum J. and Hopp, Chet J. and Boese, Carolin M. and Warren‐Smith, Emily and Chambers, Derrick and Chu, Shanna X. and Michailos, Konstantinos and Townend, John},
    title = {EQcorrscan: Repeating and Near‐Repeating Earthquake Detection and Analysis in Python},
    journal = {Seismological Research Letters},
    volume = {89},
    number = {1},
    pages = {173-181},
    year = {2017},
    month = {12},
    issn = {0895-0695},
    doi = {10.1785/0220170151},
    url = {https://doi.org/10.1785/0220170151},
    eprint = {https://pubs.geoscienceworld.org/ssa/srl/article-pdf/89/1/173/4018630/srl-2017151.1.pdf},
}

@Article{Olive2024,
author={Olive, Jean-Arthur
and Ekstr{\"o}m, G{\"o}ran
and Buck, W. Roger
and Liu, Zhonglan
and Escart{\'i}n, Javier
and Bickert, Manon},
title={Mid-ocean ridge unfaulting revealed by magmatic intrusions},
journal={Nature},
year={2024},
month={Apr},
day={01},
volume={628},
number={8009},
pages={782-787},
issn={1476-4687},
doi={10.1038/s41586-024-07247-w},
url={https://doi.org/10.1038/s41586-024-07247-w}
}

\end{document}